\documentclass{elsarticle}
\usepackage[utf8]{inputenc}
\usepackage{microtype}
\usepackage{lmodern}
\usepackage{amsmath,amsfonts,amsthm,amssymb,color}
\usepackage{hyperref}
\usepackage{bbm}
\usepackage{titlesec}
\usepackage{graphicx}

\theoremstyle{plain}

\theoremstyle{definition}

\theoremstyle{remark}

\newcommand{\corr}{\operatorname{corr}}
\newcommand{\EMA}{\operatorname{EMA}}
\newcommand{\MACD}{\operatorname{MACD}}
\newcommand{\RS}{\operatorname{RS}}
\newcommand{\RSI}{\operatorname{RSI}}

\begin{document}

\begin{frontmatter}
	\title{Advanced Strategies of Portfolio Management in the Heston Market Model}
	
	\author[1]{Jarosław Gruszka\corref{cor1}}
	\ead{jaroslaw.gruszka@pwr.edu.pl}
	
	\author[1]{Janusz Szwabi\'{n}ski}
	\ead{janusz.szwabinski@pwr.edu.pl}
	
	\cortext[cor1]{Corresponding author}
	\address[1]{Hugo Steinhaus Center, Faculty of Pure and Applied Mathematics, Wrocław University of Science and Technology}
	
	\begin{abstract}
		There is a great number of factors to take into account when building and managing an investment portfolio. It is widely believed that a proper set-up of the portfolio combined with a good, robust management strategy is the key to successful investment.  In this paper, we aim at an analysis of two aspects that may have an impact on investment performance: diversity of assets and inclusion of cash in the portfolio. We also propose two new management strategies based on the MACD and RSI factors known from technical analysis. Monte Carlo simulations within the Heston model of a market are used to perform numerical experiments.
		
	\end{abstract}



\end{frontmatter}

\section{Introduction}

Managing an investment portfolio is a vital research topic, which can be approached by means of multiple branches of science. Beginning from the pioneering work of Harry Markovitz in 1952 \cite{markowitz_portfolio_1952}, for which he was awarded a Nobel Prize, a plethora of works have been written concerning this subject. Scientists use tools developed by various disciplines in order to find the most effective methods of setting up and managing a financial portfolio. It should not come as a surprise, that a huge part of the world's portfolio management research is related to the realm of mathematics. Among the well-established ways of approaching the topic one could definitely mention theoretical mathematics, specifically --- the theory of stochastic differential equations \cite{musiela_stochastic_2010} and the theory of martingales \cite{cvitanic_hedging_1996}. Advanced statistical methods are also extremely popular, especially when it comes to the accurate estimation of risk associated with each investment opportunity \cite{jorion_international_1985, biglova_different_2004, ceria_incorporating_2006}. However, there also exist some less renowned branches of mathematics which are used in order to study portfolio dynamics and one of them is certainly the theory of fuzzy sets \cite{tanaka_portfolio_2000, carlsson_possibilistic_2002}.  Recently, with the raise of popularity of machine learning techniques, some advances in the field of portfolio management are also done, which utilise this new and powerful toolbox \cite{ban_machine_2016, paiva_decision-making_2019}. Finally, one of the branches of science which has recently contributed a lot to the research on investment portfolio dynamics is econophysics \cite{jovanovic_emergence_2013}. Looking at the vibrant financial environment from the perspective of natural sciences shed new light on the issue in general and allowed researchers to attempt seeking correlations between well-studied physical systems and the ones associated with the world of economy. 

Recently, the impact of some specific factors (e.g. transaction fees) on various portfolio management strategies was studied within the Geometric Brownian Motion (GBM) model \cite{alper_effects_2017}. The authors looked at strategies based mainly on a cyclic investor’s activity called portfolio rebalancing. We broadened their analysis in our previous paper \cite{gruszka_best_2020} as well as applied it to some real data from the Polish stock market. In this paper, we are going to extend the strategies presented therein by additional factors that may be crucial while making investment decisions:  portfolio diversification and inclusion of cash in an investment portfolio. Moreover, we are going to study the results of performing buy and sell transactions based on signals triggered by well-known technical analysis’ indicators — MACD and RSI. To this end, we introduce our own portfolio management strategies that incorporate and utilise those signals. In contrast to both of previous works mentioned \cite{alper_effects_2017,gruszka_best_2020}, here we will use the Heston model \cite{heston_closed-form_1993} of a market to generate synthetic asset data as it offers more flexibility than GBM and is more realistic as well.


The paper is structured as follows: in section \ref{sec:mod_fw} we describe and justify general assumptions for all our numerical experiments, including the Heston model. In section \ref{sec:elements} we outline all the factors of portfolio management we decided to study, we also describe there how we approached modelling them. Results of Monte Carlo experiments are presented and discussed in section \ref{sec:results}. Finally, we draw some conclusions in section \ref{sec:conclusions}.

\section{Modelling framework}
\label{sec:mod_fw}

\subsection{Basic terms and assumptions}
\label{ss:basics}

In most part of this paper, we will be operating on terms and notions which have been introduced in our previous article  \cite{gruszka_best_2020}. However, we will repeat some of the most crucial concepts here, along with their definitions, to make this paper easier to understand for a reader. 

The most basic term which we will be using throughout this entire work is a \textit{portfolio} --- a collection of investment assets held by a given party e.g. an individual or a company. In more technical terms, to have a full information of an investor's portfolio consisting of $n$ assets, at any given moment of time $t$, one needs to posses information about prices of each of those $n$ assets --- $S_1, S_2, \ldots, S_n$, as well as their amounts --- $q_1, q_2, \ldots, q_n$. Having this information, the most basic measure which can be used to evaluate portfolio performance is its value, changing over time. We call it portfolio \textit{wealth}

\begin{equation}
\label{eq:wealth}
W(t) = \sum_{i=1}^{n}{S_i(t)q_i(t)}.
\end{equation} 

In order to make it easier to compare various portfolios with each other, different measures are often introduced, which aim to make the wealth of portfolio more relative --- both in terms of the growing value of assets and the passing time. After Alper et al. \cite{alper_effects_2017} we use a measure called a \textit{growth of portfolio},

\begin{equation}
\label{eq:growth}
g(t) = \frac{\log\frac{W(t)}{W(0)}}{t}.
\end{equation}

The most important assumption that we make within all our experiments, which will be described later, is that the prices of assets do not depend on decisions a single investor makes. That means the character of trajectories $S_i$ for any $i \in \{1, \ldots, n\}$ is fully random and modelled by a dedicated stochastic process. Therefore, the only way an investor can influence performance of their portfolio is by altering quantities $q_i$ of the assets held. Another, very natural assumption is that at any point of time $t$ the investor only knows prices of each of the $n$ assets up to this moment and not further, i.e. for every $i \in \{1, \ldots, n\}$ values of $S_i(u)$ are known only for $u<t$. Finally, we assume that an investor is not able to alter the price of an asset itself by performing any market transaction. Admittedly, this is not always true in case of real markets, especially if we consider transactions performed by big market players like banks, mutual funds or hedge funds. However, for a single, individual investor, this does not need to be treated as a limitation as most of market participants operate within a range of financial means way too small to be able to influence a typical stock market.  

In our previous paper \cite{gruszka_best_2020}, multiple portfolio management strategies were studied. Among them, a \textit{passive portfolio} was the simplest one. Here, we recall its definition, as it is often used as a benchmark. In a portfolio of such kind we only decide which stocks to buy at the beginning as well as how much of them we want, but later on we do not change those quantities. Hence, for every $i \in \{1, 2, \ldots, n\}$ and for any $t \in [0, T]$ we have

\begin{equation}
\label{eq:passive_quant}
q_i(t) = q_i(0) = const.
\end{equation} 

\subsection{Need for synthetic data}

A big part of a modern research related to the methods of portfolio management focuses on testing obtained results based on real market data. Although it may seem justified and legitimate, studying financial markets only that way has some drawbacks as well. First of all, research which focuses on one particular market, stock, or even basket of stocks cannot be treated as fully universal. This is because any conclusions of such research are only fully applicable to this one particular class of assets which have been used to prove authors’ claims. In order to avoid that, we will use synthetically generated data to make experiments and hence --- we will draw more general conclusions which can be applied to any market and any stocks or sets of stocks. Using the Monte Carlo framework enabled us to generate arbitrary number of trajectories with help of stochastic processes that we considered to model our assets. Only then we performed all our numerical experiments and average the results on. Thanks to that we eradicated the bias related to picking some particular assets from existing markets. We consider our results to be more generic since they represent an average scenario of what may happen at any possible market in the world. 

\subsection{Heston model}
\label{ss:heston}

Like we mentioned in section \ref{ss:basics}, we assume that a single investor can only watch the prices of assets randomly changing on the market, not being able to influence them. As a model describing the behaviour of the assets, we chose the \textit{Heston model} \cite{heston_closed-form_1993}. This model is often used as a description of the movement of an underlying assets while pricing derivative instruments, particularly options. Heston model as well as some of its variants, are heavily studied in physics --- and are often referred to as \textit{diffusive diffusivity} models \cite{cherstvy_time_2017}. Within this model, the price of a financial asset is considered to be a stochastic process which solves the following stochastic differential equation\footnote{in the equations related to the Heston model, indices designating the $i$-th asset have been dropped for the clarity of the record}:

\begin{gather}
	\label{eq:Heston_main}
	dS(t) = \mu S(t) dt + \sqrt{v(t)} S(t) d B^S(t),\\
	dv(t) = \kappa(\theta - v(t))dt + \sigma \sqrt{v(t)} S(t) d B^v(t).
\end{gather}

As one can see, the process $S(t)$ is built upon the value of $\mu$ which is constant and the value of $v(t)$ which is a stochastic process on its own (hence --- the diffusive diffusivity term, often used by econophisicists). $\mu$ is called a drift and it represents a general tendency of an asset to grow (if $\mu >0$) or fall (if $\mu<0$). On the other hand, $v(t)$ represents the volatility of the asset and is actually modelled by a process known as CIR, originally introduced by Cox, Ingersoll and Ross to model the movement of interest rates \cite{cox_theory_1985}. One of the defining features of the CIR model is the so called \textit{mean-reversion}. The value of the process generally oscillates around a long-term average $\theta$, randomly converging to and diverging from it with the rate of $\kappa$. A random factor of severity of those oscillations is reflected in the value of $\sigma$, which hence can be called "a volatility of a volatility". Thus, the Heston model, featuring volatility of an asset changing in time, appears to reflect the behaviour of the real-life markets well. In reality, we indeed observe periods of time when prices of assets do not move significantly (practitioners often refer to such behaviour as staying in a \textit{consolidation}) but another times prices of those very same assets fluctuate strongly, achieving e.g. daily return rates, which can be orders of magnitude greater than during the peaceful times \cite{corhay_statistical_1994}. Geometric Brownian Motion, which is very often used for the purpose of simulating prices of financial assets changing in time \cite{hull_options_2018} is not able to simulate such behaviour.

The actual randomness of the prices of assets and their volatility is achieved by including Wiener processes (also known as Brownion motion processes), denoted by $B^S(t)$ and $B^v(t)$ respectively. One should allow for the possibility that those two are correlated with instantaneous correlation $\rho$

\begin{equation}
	\label{eq:Heston_rho}
	dB^S(t) dB^v(t) = \rho dt.
\end{equation}

This can also be explained from the practical point of view since there seem to be an actual correlation between prices of assets and their volatility. It is usually observed to be negative --- i.e. an increased volatility of a market usually occurs when prices drop, especially as a consequence of some kind of a market event, often related to some critical political or economical news. On the other hand, when the prices casually grow, lower market volatility can be observed \cite{schwert_stock_2011}. 

To complete the set-up of the Heston model --- initial conditions for both $S(t)$ and $v(t)$ are required

\begin{gather}
	\label{eq:Heston_ic}
	S(0) = S_0 > 0,\\
	v(0) = v_0 >0.
\end{gather}
Here, $s_0$ represents the initial price of an asset at time $t=0$, and $v_0$ is the value of the market volatility at that point of time. 

\section{Elements of portfolio construction and management strategy}
\label{sec:elements}

\subsection{Portfolio diversity}

Most investors argue that one of the most important factors of creating a successful investment portfolio is diversification \cite{lhabitant_portfolio_2017}, which is a rule behind a real-life advice \textit{not to put all one's eggs into one basket}. This essentially means not to use all available money to only buy one kind of a financial asset. Investors who do not diverse their portfolios are usually beginners \cite{goetzmann_equity_2008}, either unconscious of the risk they undertake or hoping for a one-off "golden-shot", which would allow them to quickly earn a lot of money --- a behaviour scheme arguably more similar to gambling than to responsible investing. It is worth to note that diversification is much more than just selecting more than one asset to be included in our portfolio. Stocks that we choose should belong to companies operating in different industries, be of a various size and should differ in their business model. This allows the entire portfolio to be more resilient to some bigger movements which may take place in one branch of economy, as it often happens that while some companies lose money because of some market events, others take advantage of it. 

In terms of simulations themselves, there are a few ways to create some portfolios as more diversified than others. The first method simply uses the simulation parameters. As mentioned in the previous paragraph, stocks in a well-diversified portfolio should differ from one another. In simulations, we can achieve it by varying parameters of the model generating trajectories of stock prices. For example, in the Heston model, described in section \ref{ss:heston}, one can use different values of $\mu, \kappa, \theta$ or $\sigma$ parameters to differentiate behaviour of each of the stocks. Another important concept, when it comes to the mathematical modelling of diversification, is correlation. Since companies operating in the same branches of economy usually have similar client base, not differ significantly in an internal structure and generally face almost the same external risk factors --- one can assume they will react in a similar way for all market events relevant to the given industry and hence --- their stock price plots will look quite similarly. Correlation allows us to introduce this similarity to the trajectories of stochastic processes representing those prices. Since in our case we have $n$ assets, we can speak about an entire correlation matrix $\varrho = [\rho_{i,j}]_{n\times n}$ where 

\begin{equation}
\label{eq:corr}
	\rho_{i,j} = \corr(S_i(t), S_j(t)) \text{ for any } t>0.
\end{equation}

In a well diversified portfolio, an investor should aim to maximise the number of assets which are not correlated to each other or they are correlated negatively, i.e. 

\begin{equation}
\label{eq:corr_div}
	\rho_{i,j} \leqslant 0 \text{ for some } i,j \in \{1, 2, \ldots, n\}.
\end{equation}

\subsection{Inclusion of cash in the portfolio}

Another important question investors often ask themselves is whether they should leave some money prepared for investment in form of cash and~---~if so~---~how much should they leave. Not investing all possessed money into stocks at the very first moment has some immediate advantages. It reduces the risk, as the more money remains inside a portfolio, the more stable it is, since cash, unlike stocks, does not change its value in time at all\footnote{this statement is not entirely true due to existence of inflation and a possibility of placing cash into a risk-free interest bearing deposit, although we consider neither of those in this work}. Moreover, leaving some cash aside allows an investor to react when an opportunity on a market appears, without the need to make any changes in the existing portfolio stock arrangement. On the other hand however, holding cash in a portfolio diminishes the amount of money earned form an actual investment, which seems to be especially dissatisfying in case of a very successful arrangement of the risky assets.

From the modelling perspective, including a possibility of having cash in portfolio is trivial. It is only needed to introduce a new kind of an asset, say $S_0$ and state that 

\begin{equation}
	S_0(t) = 1 \text{ for all } t \in [0, T]. 
\end{equation} 

In such a set-up, $q_0(t)$ can represent the amount of cash in a portfolio and the definition of wealth (in formula \eqref{eq:wealth}) needs to be adjusted to also take into account the "zeroth" asset 

\begin{equation}
\label{eq:wealth_updt}
W(t) = \sum_{i=0}^{n}{S_i(t)q_i(t)} = q_0(t) + \sum_{i=1}^{n}{S_i(t)q_i(t)}.
\end{equation} 

\subsection{MACD trading indicator}

The last factor discussed in this paper is trading indicators. Throughout the years, investors have been attempting to find some mathematical tools which would allow them to "predict the future" or, at least, to help them find the correct moment to buy or sell a stock. This resulted in creation of a wide set of indicators and markers for this purpose. A branch of trading activities which studies usefulness of those markers is called \textit{technical analysis} \cite{farias_nazario_literature_2017}. One of the most widespread indicators, well known among all investors using technical analysis (often referred to as traders, due to usually very short time horizons of their investments) is MACD --- Moving Average Convergence Divergence \cite{appel_technical_2005}.

MACD was invented by Gerard Appel in 1979. It is based on three time series which are derived from the asset price process by means of a transformation very commonly used in technical analysis --- EMA --- exponential moving average. EMA, as the name itself suggests, is a kind of a moving average, but with exponentially decreasing weights of factors more distant in time from the current one. It can easily be described by a recursive formula

\begin{equation}
\label{eq:ema}
	\EMA_{X, p}(t) = \begin{cases}
	X(t), &\text{ for } t = 0\\
	\alpha X(t) + (1-\alpha) \EMA_{X,p}(t-\Delta t), &\text{ for } t > 0 
	\end{cases}
\end{equation}
where $\alpha = \frac{2}{p+1}$. EMA is time dependent and has two parameters. One is the base process $X$. In case of MACD --- this process is simply the stock price of an $i$-th asset $S_i$. 
The second parameter $p$ is called the lag and it is largely responsible for the weight of the past values taken to the average. The bigger $p$, the bigger is the weight of older values of the underlying process in the final result, which has an effect in a smoother EMA curve.

EMA as a discrete operator, requires time discretisation. In order to use eq. \eqref{eq:ema}, one needs to fix a small time interval of a length $\Delta t$ and EMA will only be available at the time points $t = k\Delta t, k\in \{1,2,\ldots, K-1\}$ such that $(K-1)\Delta t \leqslant T$ and $K\Delta t > T$. 

MACD indicator makes its predictions based on the difference between two time series. The first is the MACD line, i.e. a line obtained by subtracting two EMAs of different lags. 

\begin{equation}
\label{eq:macd}
\MACD_{i, p,q}(t) = \EMA_{S_i, p}(t)-\EMA_{S_i, q}(t)
\end{equation}
for $p<q$. EMA related to the lag parameter $p$ is called the fast line whereas the one related to the parameter $q$ is called the slow line. Buy and sell signals are generated by places where the MACD line crosses what is called, the signal line --- another EMA, with a new lag parameter $s<p$. In other words, we can introduce the final indicator line as

\begin{equation}
F_{i,p,q,s}(t) = \EMA_{S_i, s}(t)-\MACD_{i,p,q}(t).
\end{equation}

Whenever this line changes its value from negative to positive --- MACD gives a "buy" signal. Contrarily, when this line drops form positive values to the negative ones --- we obtain a "sell" signal. Hence we, can introduce the buying and selling indicators:

\begin{equation}
\label{eq:macd_buy_ind}
\mathbbm{1}_{i,p,q,s}^{+}(t) = \begin{cases}
1, &\text{ if } F_{i,p,q,s}(t-\Delta t) < 0 \land F_{i,p,q,s}(t) > 0\\
0, &\text{ otherwise.} 
\end{cases}
\end{equation}

\begin{equation}
\label{eq:macd_sell_ind}
\mathbbm{1}_{i,p,q,s}^{-}(t) = \begin{cases}
1, &\text{ if } F_{i,p,q,s}(t-\Delta t) > 0 \land F_{i,p,q,s}(t) < 0\\
0, &\text{ otherwise.} 
\end{cases}
\end{equation}

Since the MACD indicator only tells \textit{when} to buy or sell a particular asset, but not \textit{how much}, we created our own strategy for that purpose. Let us therefore introduce factors $\psi$ and $\phi$, such that $\psi, \phi \in [0,1]$. They can be picked arbitrarily by any investor, as they are meant to represent their trust in selling and buying signals, generated by the MACD indicator. 

$\psi$ is called the sell factor. Whenever MACD generates a sell signal for a given stock, a $\psi$ part of the amount of this asset is sold and the money from selling is converted into portfolio cash,

\begin{align}
\label{eq:macd_q_update1}
q_i'(t) &= q_i(t-\Delta t) (1-\psi\mathbbm{1}_{i,p,q,s}^{-}(t)),\\
\label{eq:macd_q0_update1}
q_0'(t) &= q_0(t-\Delta t) + \sum_{i=1}^{n}{S_i(t)q_i(t-\Delta t)\psi\mathbbm{1}_{i,p,q,s}^{-}(t)}.
\end{align}

Similarly, $\phi$ is called the buy factor. This time however, it is used to decide, what part of available cash (including new portion obtained from selling some stocks) will be used to buy new stocks

\begin{equation}
\label{eq:macd_cash_to_buy}
	c(t) = \phi q_0'(t).
\end{equation}

The fraction of portfolio cash $c(t)$ is then used to buy assets indicated by a buy signal

\begin{align}
\label{eq:macd_q_update_final}
q_i(t) &= q_i'(t) +\frac{c(t)}{S_i(t)}\left(\sum_{i=1}^{n}{\mathbbm{1}_{i,p,q,s}^{+}(t)}\right)^{-1}\mathbbm{1}_{i,p,q,s}^{+}(t),\\
\label{eq:macd_q0_update_final}
q_0(t) &= q_0'(t)-c(t).
\end{align}

It can easily be shown that this strategy is self-financing and the amount of cash in portfolio will never drop below zero (see \ref{sec:appendix} for a proof).

\subsection{RSI trading indicator}

An other, commonly used trading indicator is called RSI --- the relative Strength Index. It was discovered by J. Welles Wilder Jr. in 1978 \cite{wilder_new_1978}. Values of the index can only be in the interval $[0, 100]$ and the marker can be used to identify when an instrument is oversold (index' value below certain level, usually 30 --- a signal for buying) or when it is overbought (index' value above certain level, usually 70 --- a signal for selling). In order to calculate the value of RSI, again prices need to be discretised with respect to the time period $\Delta t$ and then differences between subsequent prices need to be calculated

\begin{equation}
\label{eq:rsi_diff}
D_i(t) = \begin{cases}
0, &\text{ for } t = 0,\\
S_i(t) - S_i(t-\Delta t), &\text{ for } t > 0.
\end{cases}
\end{equation}

Next, positive and negative differences are sorted out from each other

\begin{equation}
D^+_i(t) = \begin{cases}
D(t), &\text{ where } D(t) > 0\\
0, &\text{ where } D(t) \leqslant 0. 
\end{cases}
\end{equation}

\begin{equation}
D^-_i(t) = \begin{cases}
0, &\text{ where } D(t) \geqslant 0\\
D(t), &\text{ where } D(t) < 0. 
\end{cases}
\end{equation}

Those two time series are transformed through EMA and additionally averaged by a classical, arithmetic mean. This gives two crucial coefficients for calculating the value of RSI --- $a_i$ and $b_i$ 

\begin{equation}
a_i(t) = \frac{1}{\#\{k: k\in \mathbb{N}, k\Delta t \leqslant t\}}\sum_{\{k: k\in \mathbb{N}, k\Delta t \leqslant t\}}\EMA_{D^+_i,s}(k\Delta t),
\end{equation}

\begin{equation}
b_i(t) = \frac{1}{\#\{k: k\in \mathbb{N}, k\Delta t \leqslant t\}}\sum_{\{k: k\in \mathbb{N}, k\Delta t \leqslant t\}}\EMA_{D^-_i,s}(k\Delta t).
\end{equation}

The ratio of $a_i$ to $b_i$ is called \textit{relative strength}

\begin{equation}
\RS_i(t) = \frac{a_i}{b_i}.
\end{equation}

Having the relative strength, to calculate RSI one should only rescale $\RS_i$, so that all values are between 0 and 100 using a simple formula

\begin{equation}
\RSI_i(t) = 100 - \frac{100}{1+RS_i(t)}.
\end{equation}

In order to construct the actual strategy out of the values of RSI, we can construct dedicated indicators, similar to the ones that have been proposed for the MACD in eqs. \eqref{eq:macd_buy_ind} and \eqref{eq:macd_sell_ind}. This time however arbitrary levels of overbuying and overselling need to be fixed additionally. Let us denote them by $d^+$ and $d^-$ respectively. As mentioned above, practitioners usually stick to $d^+ = 30$ and $d^- = 70$. Having that fixed, we can define the following indicators for RSI strategy

\begin{equation}
\label{eq:rsi_buy_ind}
\mathbbm{1}_{i,d^+}^{+}(t) = \begin{cases}
1, &\text{ if } \RSI_i(t) < d^+,\\
0, &\text{ otherwise,} 
\end{cases}
\end{equation}

\begin{equation}
\label{eq:rsi_sell_ind}
\mathbbm{1}_{i,d^-}^{-}(t) = \begin{cases}
1, &\text{ if } \RSI_i(t) > d^-,\\
0, &\text{ otherwise.} 
\end{cases}
\end{equation}

With those indicators, we can construct the exact same strategy as in case of MACD (reacting for signals, as described by equations \eqref{eq:macd_q_update1} -- \eqref{eq:macd_q0_update_final}) simply by replacing MACD-related buy and sell indicators by the newly defined RSI-based ones. 

\section{Results}
\label{sec:results}


In this section, we present the results of our Monte Carlo experiments illustrating the impact of the factors introduced in the previous sections on the portfolio performance.  In each experiment, a number of portfolios were generated independently by sweeping the possible values of one particular factor and keeping the others fixed. Results were averaged over a certain number of independent runs, varying depending on an experiment\footnote{each plot, along with other simulation parameters, has an $MCt$ indicated in its caption --- this is exactly the number of \textbf{M}onte \textbf{C}arlo independent \textbf{t}rials which were run to obtain given result}. In all figures presented below, the growth of portfolio \eqref{eq:growth} is used as a measure of portfolio performance.

\begin{figure}[h]
	\centering
	\includegraphics[width=\textwidth]{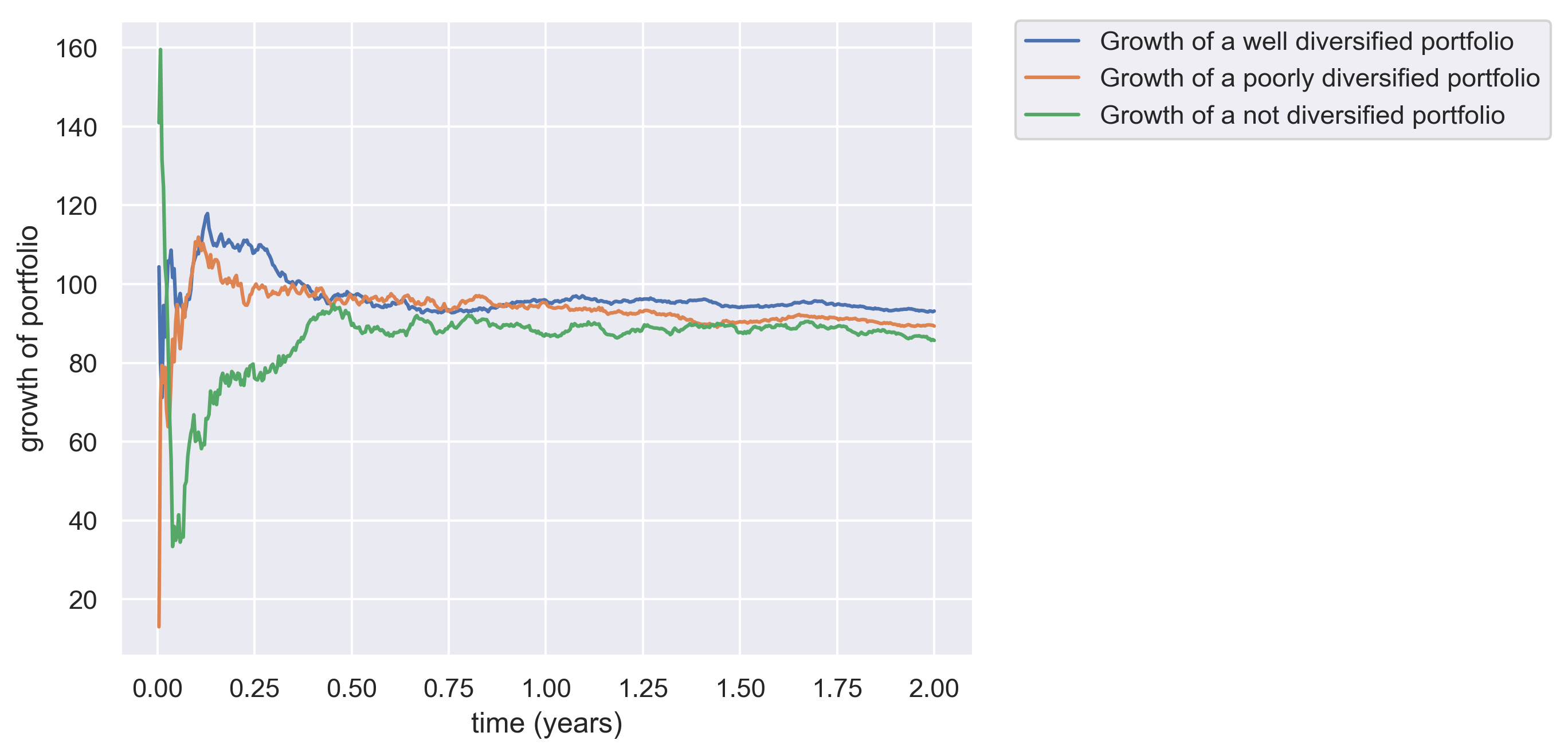}
	\caption{(Colour online) Portfolio growth in time for various types of portfolios with different level of diversification. Simulation parameters: $T = 2, \Delta t = 2^{-8}, s_0 = 100, \mu = 0.1, v_0 = 0.035, \sigma = 0.5, \rho = -0.66, MCt=1000$.}
	\label{fig:portfolioGrowth_time_diversification}
\end{figure}

In Fig. \ref{fig:portfolioGrowth_time_diversification} the importance of diversification is shown. We compared 3 types of passively run portfolios. The \textit{well diversified} portfolio consisted of 3 uncorrelated assets. Their trajectories were simulated with slightly different values of parameters $\kappa$ and $\theta$. That made the variance process of each asset have a different stochastic character. The \textit{poorly diversified} portfolio also consisted of three assets, although this time they were all positively correlated. Moreover, this time simulation parameters for all of the assets where exactly the same. The third portfolio was \textit{not diversified} and it only had one asset in it, for which prices were generated with identical parameters as in case of a poorly diversified one. As one can see, a not diversified portfolio performs definitely the worst for most of the time, whereas the well diversified and poorly diversified portfolios seem to initially compete, but the well diversified one eventually turns out to be the best in case of a 2-year time horizon.

\begin{figure}[h]
	\centering
	\includegraphics[width=\textwidth]{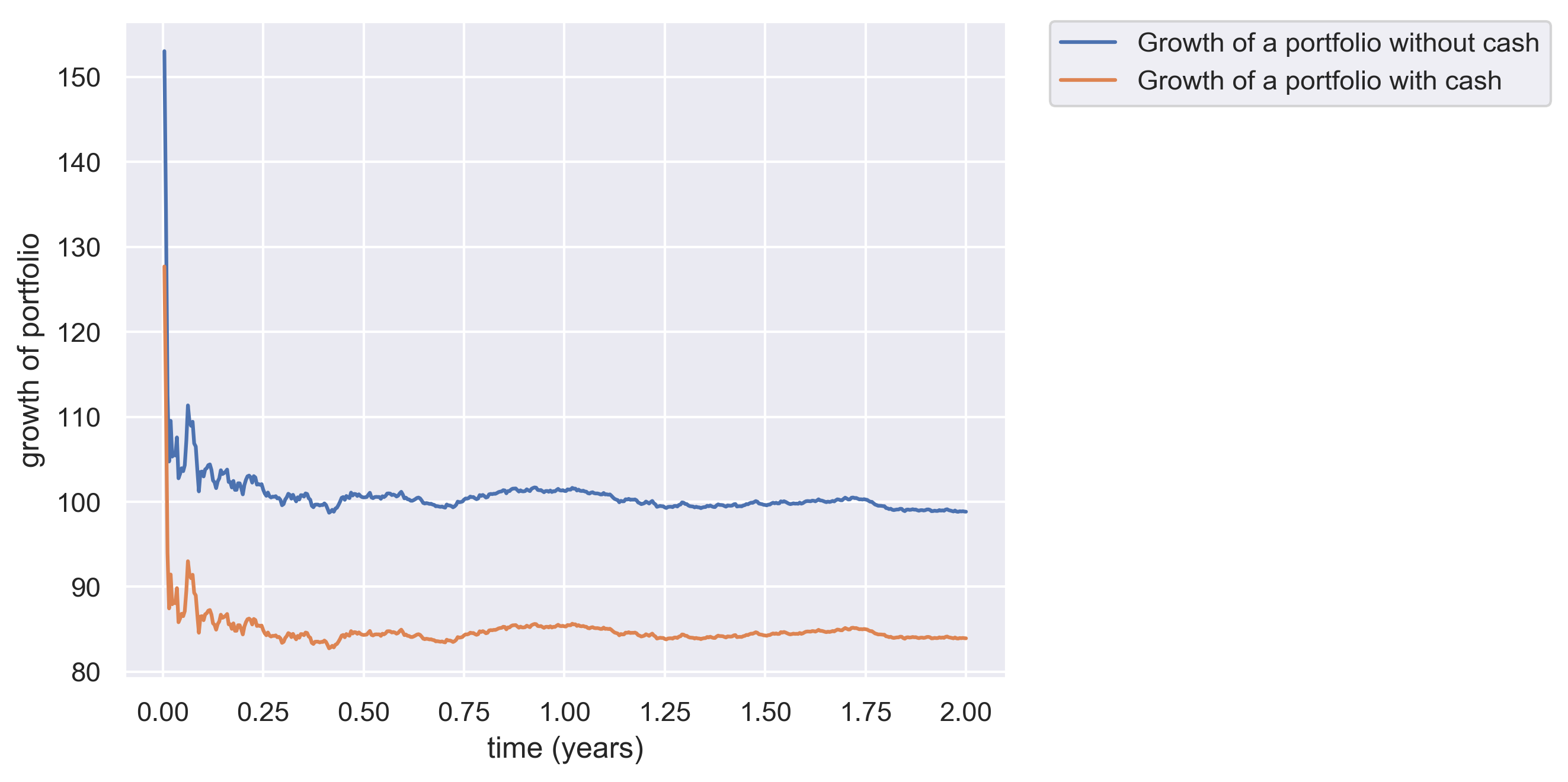}
	\caption{(Colour online) Portfolio growth in time for a passive portfolio and a portfolio with a cash contribution (approx. 28\%) for $\mu = 0.1$. Other simulation parameters: $T = 2, \Delta t = 2^{-8}, n=5, s_0 = 100, v_0 =0.025, \kappa = 1.2,  \theta = 0.05, \sigma = 0.5, \rho = -0.66, MCt=1000$.}
	\label{fig:portfolioGrowth_time_cashInclusion_bigMu}
\end{figure}

\begin{figure}[h]
	\centering
	\includegraphics[width=\textwidth]{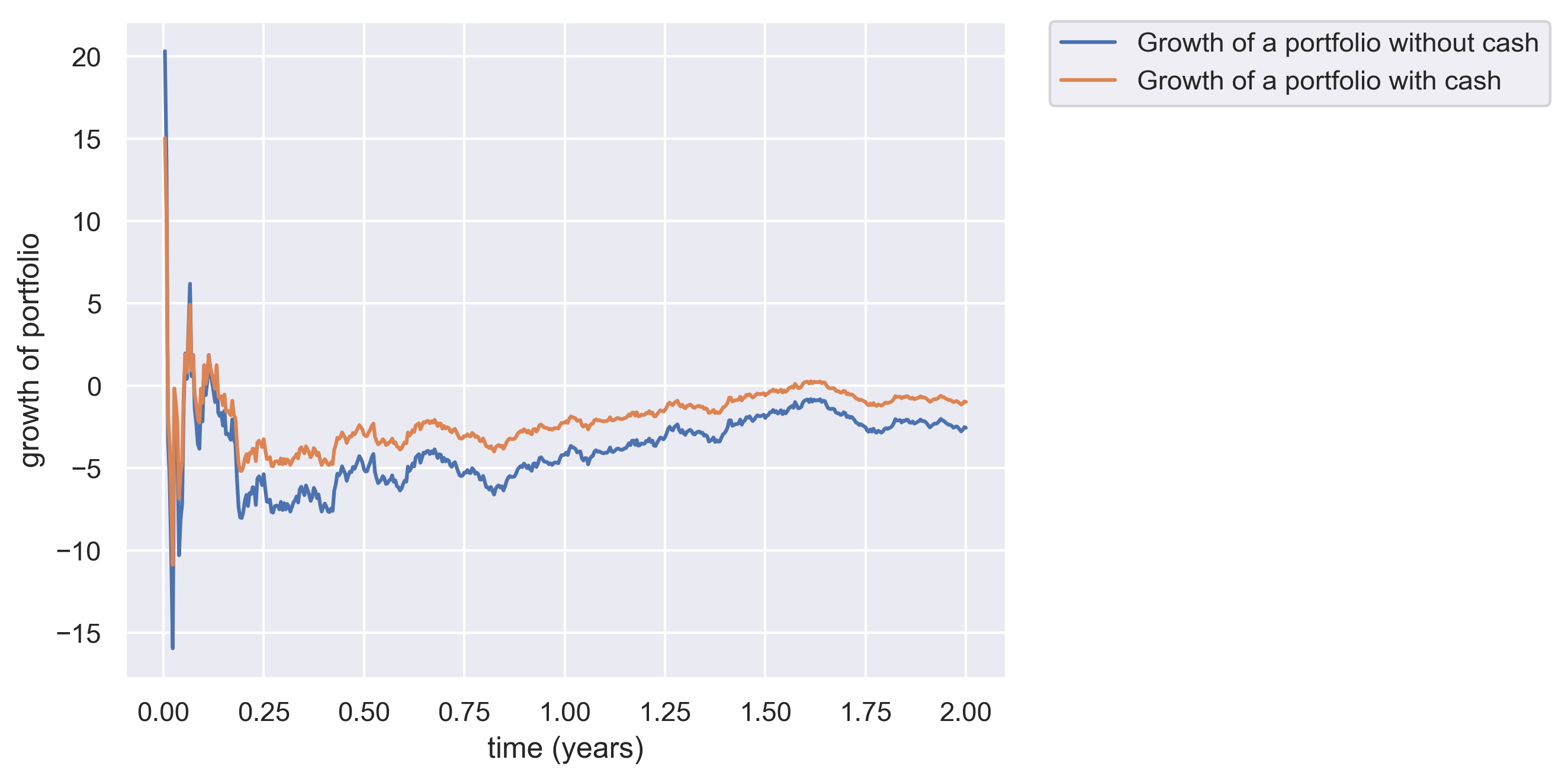}
	\caption{(Colour online) Portfolio growth in time for a passive portfolio and a portfolio with a cash contribution (approx. 28\%) for $\mu = 0.001$. Simulation parameters: $T = 2, \Delta t = 2^{-8}, n=5, s_0 = 100, v_0 =0.025, \kappa = 1.2,  \theta = 0.05, \sigma = 0.5, \rho = -0.66, MCt=1000$.}
	\label{fig:portfolioGrowth_time_cashInclusion_smallMu}
\end{figure}

\begin{figure}[h]
	\centering
	\includegraphics[width=\textwidth]{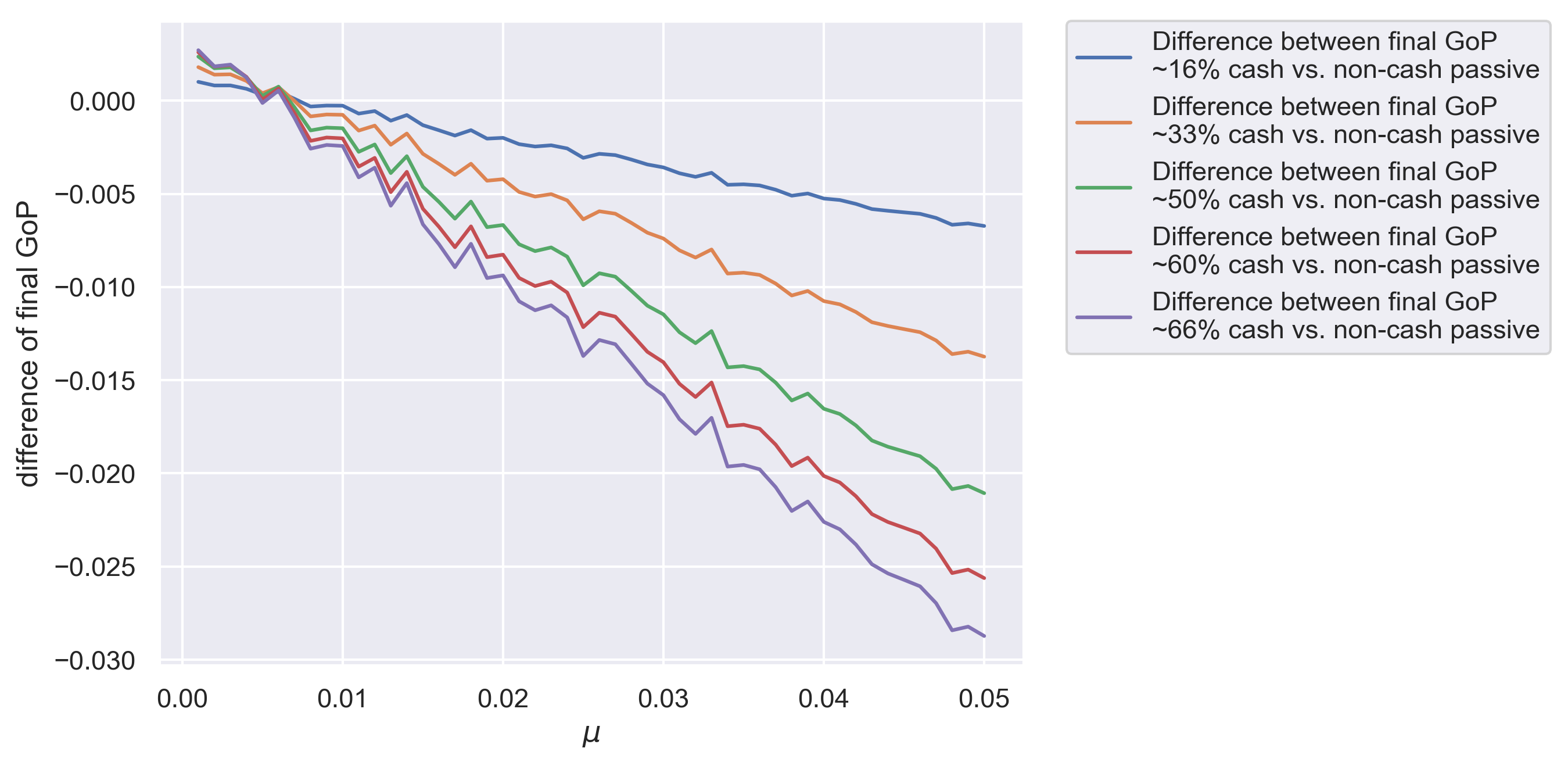}
	\caption{(Colour online) Impact of the amount of cash stored in the portfolio for various values of the drift parameter $\mu$. Simulation parameters: $T = 2, \Delta t = 2^{-8}, n=5, s_0 = 100, v_0 =0.025, \kappa = 1.2,  \theta = 0.05, \sigma = 0.5, \rho = -0.66, MCt=5000$.}
	\label{fig:differenceFinalGoP_Mu_cashInclusion}
\end{figure}

Figures \ref{fig:portfolioGrowth_time_cashInclusion_bigMu}--\ref{fig:differenceFinalGoP_Mu_cashInclusion} illustrate the impact of storing cash in an investment portfolio. The intuition here would be that a portfolio with left-away money will perform worse compared to an analogous portfolio with all money resources invested in stocks.  Rather interestingly however, it turns out it might not always be the case. We found out that stocks' drift is the decisive parameter here. In Fig. \ref{fig:portfolioGrowth_time_cashInclusion_bigMu} two passive portfolios are compared --- one with some share of cash, an one without it. We used a relatively big value of the $\mu$ parameter, meaning stock prices had a strong tendency to grow over time. In this case our expectation seems to be true --- the non-cash portfolio performs much worse. If we however change $\mu$ to a much smaller value and hence --- purge the stocks of their growth potential --- we observe quite the opposite result, pictured in Fig. \ref{fig:portfolioGrowth_time_cashInclusion_smallMu}. In such case results are better for a conservative portfolio in which cash was not entirely spent. Those two results are special cases of the third analysis focusing on the aspect of cash in the investment portfolio more generally, presented in Fig. \ref{fig:differenceFinalGoP_Mu_cashInclusion}. The plot shows the difference between the final (measured at $t=T$) growth of a classical passive portfolio, with all money invested and an analogous portfolio, having some initial share of cash --- varying for each of the few lines on the plot. Positive values of the curves mean that a portfolio with cash is better, negative ones --- quite the opposite --- an all-in portfolio gives a better outcome. Entire analysis has been performed against the $\mu$ parameter. We can clearly see it confirms that for small values of the drift a cash-featuring portfolios lead to better results. What we can also read from the plot is that the share of cash matters a lot. A portfolio consisting mostly of cash will lead to much worse results in case if stocks have a strong tendency to grow, but it will also give best results in case assets have lower growth potential.

\begin{figure}[h]
	\centering
	\includegraphics[width=\textwidth]{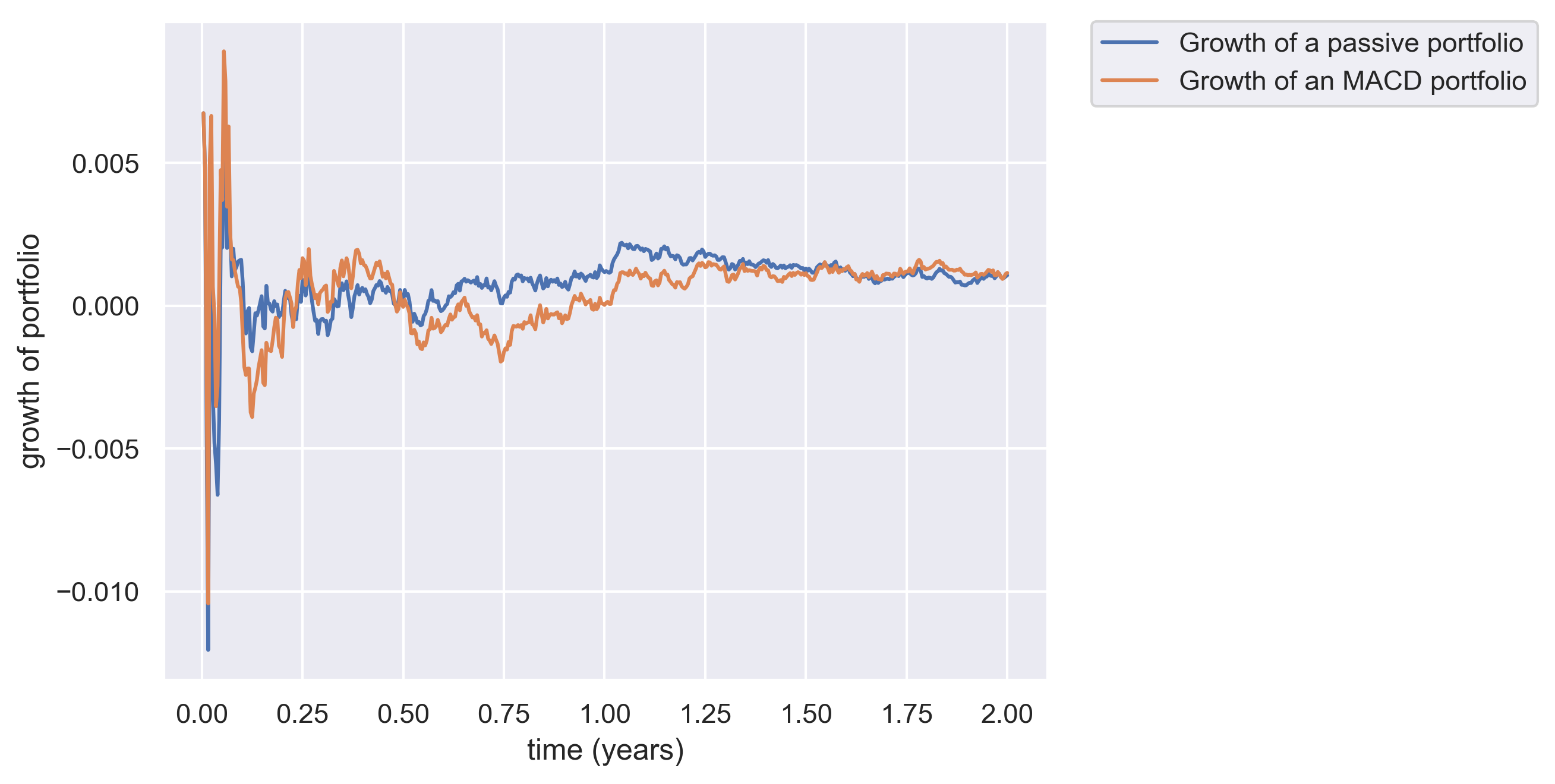}
	\caption{(Colour online) Portfolio growth in time for a passive portfolio and a MACD portfolio for $\mu = 0.005$. Other simulation parameters: $T = 2, \Delta t = 2^{-8}, n=10, s_0 = 100, v_0 =0.025, \kappa = 1.2,  \theta = 0.05, \sigma = 0.5, \rho = -0.66, MCt=1000$.}
	\label{fig:portfolioGrowth_time_MACD_smallMu}
\end{figure}

It turns out that the drift is a critical parameter not only when it comes to the cash inclusion, but it also has a huge impact on the effectiveness of strategies based on using technical analysis indicators like MACD or RSI. In Fig. \ref{fig:portfolioGrowth_time_MACD_smallMu}, a comparison between a passive and MACD-driven portfolio has been presented, for a relatively small value of $\mu$. As one can see, it is not obvious to assess which of those two perform better in this case. If we however increase $\mu$ --- it becomes clear that portfolios managed by a strategy using MACD as an indicator outperforms a simple buy-and-hold strategy. 

\begin{figure}[h]
	\centering
	\includegraphics[width=\textwidth]{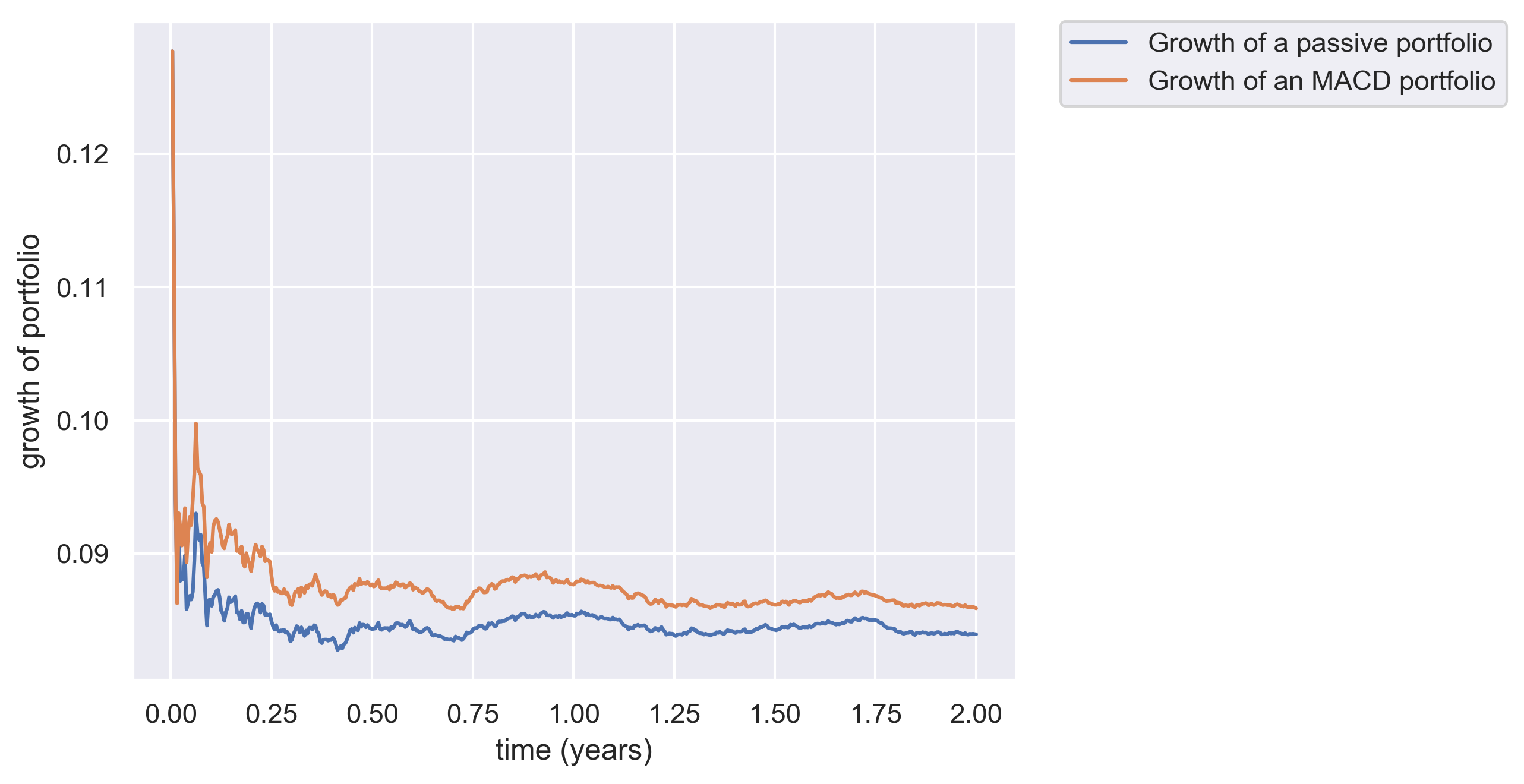}
	\caption{(Colour online) Portfolio growth in time for a passive portfolio and a MACD portfolio for $\mu = 0.1$. Other simulation parameters: $T = 2, \Delta t = 2^{-8}, n=10, s_0 = 100, v_0 =0.025, \kappa = 1.2,  \theta = 0.05, \sigma = 0.5, \rho = -0.66, MCt=1000$.}
	\label{fig:portfolioGrowth_time_MACD_bigMu}
\end{figure}

\begin{figure}[h]
	\centering
	\includegraphics[width=\textwidth]{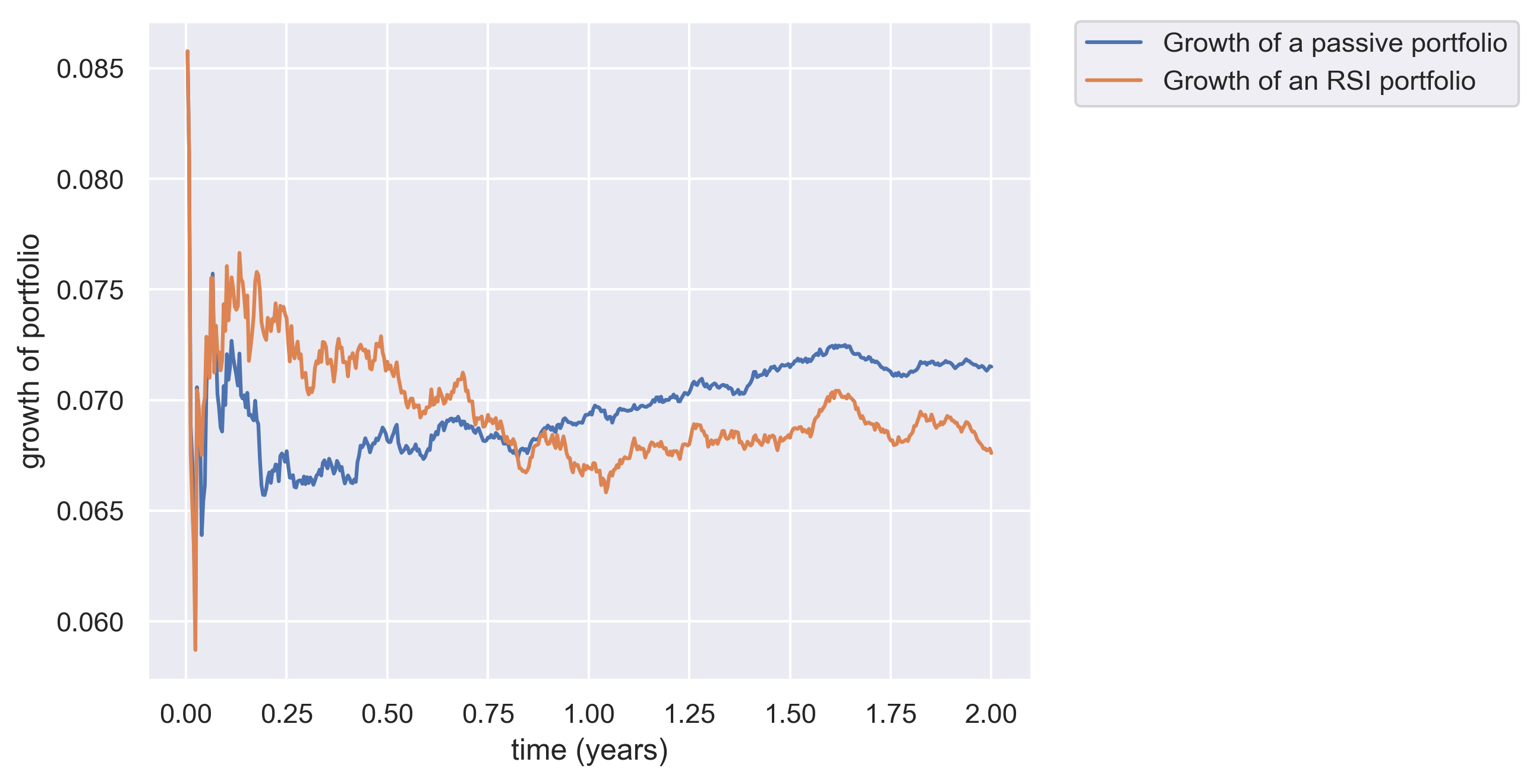}
	\caption{(Colour online) Portfolio growth in time for a passive portfolio and a RSI portfolio for $\mu = 0.1$. Other simulation parameters: $T = 2, \Delta t = 2^{-8}, n=5, s_0 = 100, v_0 =0.025, \kappa = 1.2,  \theta = 0.05, \sigma = 0.5, \rho = -0.66, MCt=1000$.}
	\label{fig:portfolioGrowth_time_RSI_smallMu}
\end{figure}

\begin{figure}[h]
	\centering
	\includegraphics[width=\textwidth]{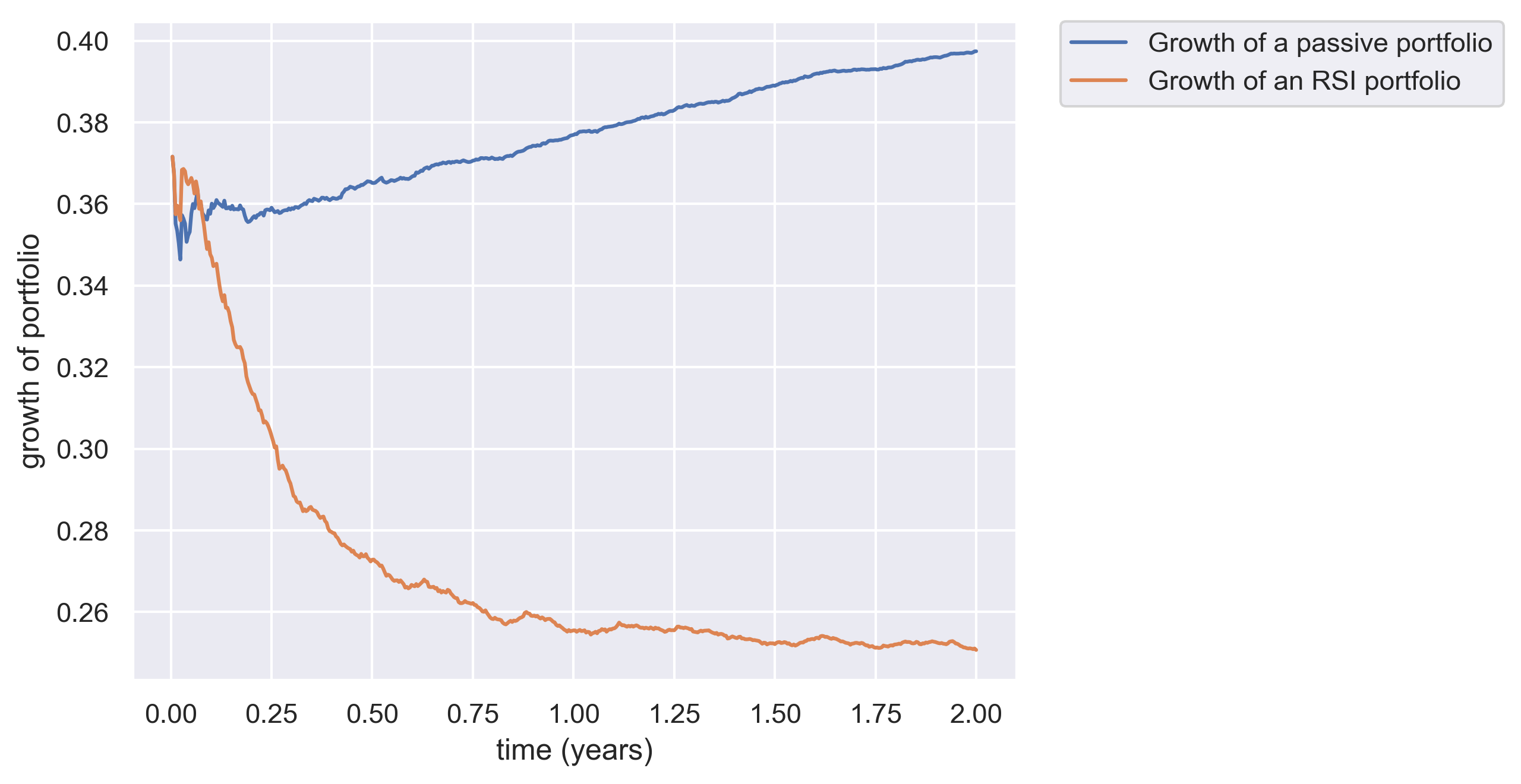}
	\caption{(Colour online) Portfolio growth in time for a passive portfolio and a RSI portfolio for $\mu = 0.5$. Other simulation parameters: $T = 2, \Delta t = 2^{-8}, n=5, s_0 = 100, v_0 =0.025, \kappa = 1.2,  \theta = 0.05, \sigma = 0.5, \rho = -0.66, MCt=1000$.}
	\label{fig:portfolioGrowth_time_RSI_bigMu}
\end{figure}

One could expect that it would be the case for strategies utilising various indicators, that the bigger the drift, the better the results. It turns out that it is very much dependent on the actual indicator being used. To demonstrate this, one can compare Fig.\ref{fig:portfolioGrowth_time_RSI_smallMu} and \ref{fig:portfolioGrowth_time_RSI_bigMu} which both present the comparison between a passive and a RSI-driven portfolio --- the former for a small value of $\mu$ and the latter --- for a bigger one. We can see that the tendency is exactly the opposite in this case. When the stocks behave more indecisively, the RSI-driven portfolio competes with the passive one on a level playing field and although the passive one wins eventually --- a shorter (e.g. a $\frac{1}{4}-$ to a $\frac{1}{2}$-year) portfolio would turn out to perform better if driven by RSI. However, when assets grow fast --- an RSI-based portfolio performs much worse than the passive one. 

\begin{figure}[h]
	\centering
	\includegraphics[width=\textwidth]{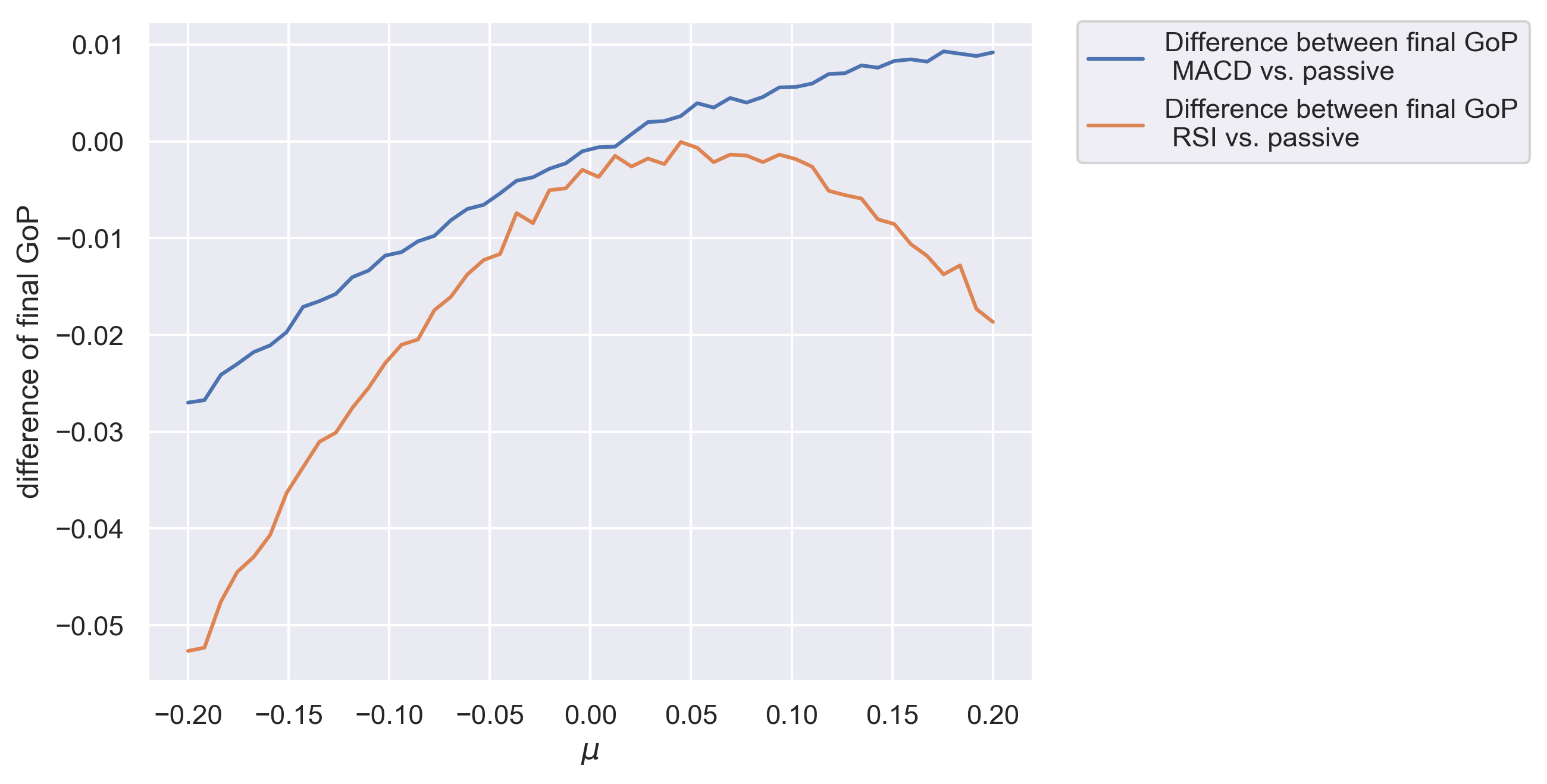}
	\caption{(Colour online) Difference between MACD and RSI portfolios and a passive portfolio for various values of the drift parameter $\mu$. Simulation parameters : $T = 2, \Delta t = 2^{-8}, n=5, s_0 = 100, v_0 =0.025, \kappa = 1.2,  \theta = 0.05, \sigma = 0.5, \rho = -0.66, MCt=5000$.}
	\label{fig:differenceFinalGoP_Mu_MACD_RSI}
\end{figure}

In order to obtain even more general results, similarly to what we have done for the cash inclusion experiment, also here we have drawn the plot of final (at $t=T$) differences between actively managed and passive portfolios in dependence of the drift parameter $\mu$ shown in Fig. \ref{fig:differenceFinalGoP_Mu_MACD_RSI}. As we can see, for the range of $\mu$ that has been used, MACD performs best for big values of the drift, whereas RSI turns out to be useful around the value of $\mu=0$ and even then it does not really outperform a passive portfolio. 

\section{Conclusions}
\label{sec:conclusions}

In the article we focused on a few important aspects of a portfolio that can be considered a well-managed one and we studied their actual importance, based on the synthetic data, simulated according to the Heston model, in a Monte Carlo-type experiments. First of those aspects is portfolio diversification. We managed to numerically show that a simple passive portfolio which consists of independent assets, all having different inner characteristics (modelled by different simulation parameters) lead to better results compared to portfolios with correlated assets of similar character. An utterly not-diversified portfolio, meaning the one with only one asset, performed definitely worst than the two diversified ones, which makes the importance of diversification even more evident. The other factor that we considered in our research was the share of cash in a portfolio. Rather surprisingly, we found out that keeping cash in a portfolio may be beneficial and that the main factor which needs to be considered here is the assets' drift. Our results reveal that if the stocks have a big tendency to grow --- leaving cash in one's portfolio deteriorates portfolio's performance. However, if this is not the case and the stocks do not represent a very significant drift --- it is exactly the opposite. The more money one keeps uninvested, the better results the portfolio attains. Finally, for our last experiment, we wanted to examine the effectiveness of popular trading indicators, commonly used by investors utilising technical analysis. We chose two of them  ---  MACD and RSI. For the purpose of testing we designed special portfolio management strategies, in which transactions were performed when an indicator sent a buy or sell signal. The comparison between those strategies and a simple passive one was very much dependent on the drift parameter again, similarly as in case of cash inclusion. Moreover, results obtained for MACD were very different from those for RSI which can serve as a confirmation that different trading indicators should be used for different market conditions.

There are really wide perspectives of continuing the above research. Including prices jumps, which can be seen in the real markets, is certainly one of them. The other one can be comparing various portfolio performance measures, as the growth of portfolio, which we used is not universally acclaimed to be the best one. Finally, it would be very interesting to find new management strategies, which would be more successful than the ones based on MACD and RSI and possibly less dependent on the assets drift. Those and multiple other questions which arose during writing this work definitely constitute a great incentive for continuing the research in the field.

\section*{Acknowledgement}

This work was supported by the Polish Ministry of Science and Higher Education (MNiSW) core funding for statutory R\&D activities.

\bibliographystyle{elsarticle-num}
\bibliography{Bibliography}

\appendix
\section{}
\label{sec:appendix}

Below we present the proof of the fact that the strategy described by equations \eqref{eq:macd_q_update1} -- \eqref{eq:macd_q0_update_final} is self financing, i.e. at any point of time (except $t=0$ and we set up our portfolio) only resources available in portfolio are used and no value is added to or withdrawn from portfolio. 

We start by fixing the initial amount of cash in the portfolio $q_0(0) > 0$ and initial amounts of assets that we buy at the beginning $q_i(0)>0, i\in \{1, 2, \ldots, n\}$. 

Let us now look into the movements which happen at an arbitrary point of time $t$. On one hand, we know that if we consider having cash in our portfolio, the wealth of it should be expressed by formula \eqref{eq:wealth_updt}. This is what we call the \textit{actual wealth},

\begin{equation}
\label{eq:wealth_ext}
	W(t) = q_0(t) + \sum_{i=1}^{n}{S_i(t)q_i(t)}.
\end{equation}

On the other hand however, we know that before we do any buy or sell transaction at time $t$, the value of our portfolio arises from the amount of assets we had in the previous step. This is what we call a \textit{temporary wealth}   

\begin{equation}
\label{eq:wealth_temp}
	W^{temp}(t) = q_0(t-\Delta t) + \sum_{i=1}^{n}{S_i(t)q_i(t-\Delta t)}.
\end{equation}

Proving that the strategy is self financing means proving that those two wealths are equal, i.e. we only use financial resources that we have available because of our previous investment decisions,

\begin{equation}
\label{eq:wealth_equal}
W^{temp}(t) = W(t).
\end{equation}
Using equations \eqref{eq:macd_q_update_final} and \eqref{eq:macd_q0_update_final} we have 

\begin{equation*}
\begin{split}
W(t) &= q_0(t) + \sum_{i=1}^{n}{S_i(t)q_i(t)} = \\
&= q_0'(t)-c(t) + \sum_{i=1}^{n}{S_i(t)\left(q_i'(t) +\frac{c(t)}{S_i(t)}\left(\sum_{j=1}^{n}{\mathbbm{1}_{j,p,q,s}^{+}(t)}\right)^{-1}\mathbbm{1}_{i,p,q,s}^{+}(t)\right)} =  \\
&= q_0'(t)-c(t) + \sum_{i=1}^{n}{S_i(t)q_i'(t)} + \sum_{i=1}^{n}{S_i(t)\frac{c(t)}{S_i(t)}\left(\sum_{j=1}^{n}{\mathbbm{1}_{j,p,q,s}^{+}(t)}\right)^{-1}\mathbbm{1}_{i,p,q,s}^{+}(t)} = \\
&= q_0'(t)-c(t) + \sum_{i=1}^{n}{S_i(t)q_i'(t)} + \sum_{i=1}^{n}{c(t)\mathbbm{1}_{i,p,q,s}^{+}(t)\left(\sum_{j=1}^{n}{\mathbbm{1}_{j,p,q,s}^{+}(t)}\right)^{-1}} = \\
&= q_0'(t)-c(t) + \sum_{i=1}^{n}{S_i(t)q_i'(t)} + 
c(t) \left(\sum_{j=1}^{n}{\mathbbm{1}_{j,p,q,s}^{+}(t)}\right)^{-1}\sum_{i=1}^{n}{\mathbbm{1}_{i,p,q,s}^{+}(t)} = \\
&=q_0'(t)-c(t) + \sum_{i=1}^{n}{S_i(t)q_i'(t)} + c(t) = q_0'(t) + \sum_{i=1}^{n}{S_i(t)q_i'(t)}
\end{split}
\end{equation*}
Now, using equations \eqref{eq:macd_q_update1} and \eqref{eq:macd_q0_update1} we obtain

\begin{equation*}
\begin{split}
W(t) &= q_0'(t) + \sum_{i=1}^{n}{S_i(t)q_i'(t)} = \\
&= q_0(t-\Delta t) + \sum_{i=1}^{n}{S_i(t)q_i(t-\Delta t)\psi\mathbbm{1}_{i,p,q,s}^{-}(t)} + \\
&+ \sum_{i=1}^{n}{S_i(t)q_i(t-\Delta t) (1-\psi\mathbbm{1}_{i,p,q,s}^{-}(t))} = \\
&=q_0(t-\Delta t) + \sum_{i=1}^{n}{S_i(t)q_i(t-\Delta t)\psi\mathbbm{1}_{i,p,q,s}^{-}(t)} + \\
&+\sum_{i=1}^{n}{S_i(t)q_i(t-\Delta t)} - \sum_{i=1}^{n}{S_i(t)q_i(t-\Delta t)\psi\mathbbm{1}_{i,p,q,s}^{-}(t)} = \\
&= q_0(t-\Delta t) + \sum_{i=1}^{n}{S_i(t)q_i(t-\Delta t)} = W^{temp}(t)
\end{split}
\end{equation*}

Thus, since $W(t) = W^{temp}(t)$, the strategy is indeed self-financing. 

It can also be proven that the amount of cash after each step will never be negative. More precisely, we will prove that if an amount of cash brought from the previous time step is non-negative, it will remain such in the next step. That, together with an assumption that the initial amount of cash at time $t=0$ is non-negative proves it will always be that way. 

The ultimate amount of cash at time $t$ is given by equation \eqref{eq:macd_q0_update_final}

\begin{equation*}
q_0(t) = q_0'(t)-c(t).
\end{equation*}
Plugging equation \eqref{eq:macd_cash_to_buy} into the above one we get

\begin{equation*}
q_0(t) = q_0'(t)- \phi q_0'(t) = q_0'(t) (1-\phi).
\end{equation*}
Using equation \eqref{eq:macd_q0_update1} what we obtain is

\begin{multline*}
q_0(t) = \left(q_0(t-\Delta t) + \sum_{i=1}^{n}{S_i(t)q_i(t-\Delta t)\psi\mathbbm{1}_{i,p,q,s}^{-}(t)}\right) (1-\phi) = \\
= \left(q_0(t-\Delta t) + \psi\sum_{i=1}^{n}{S_i(t)q_i(t-\Delta t)\mathbbm{1}_{i,p,q,s}^{-}(t)}\right) (1-\phi).
\end{multline*}

\newpage

Now, since $q_0(t-\Delta t) \geqslant 0$ (by assumption), as well as $\psi \in [0,1]$, $S_i(t) > 0$, $\mathbbm{1}_{i,p,q,s}^{-}(t) \geqslant 0$ and $q_i(t-\Delta t) \geqslant 0$ (by their respective definitions), the entire expression $q_0(t-\Delta t) + \psi\sum_{i=1}^{n}{S_i(t)q_i(t-\Delta t)\mathbbm{1}_{i,p,q,s}^{-}(t)}$ must be non-negative. Also, since $\phi \in [0,1]$, the expression $(1-\phi) \geqslant 0$. Hence, multiplication of those two expressions must also be non-negative and thus

\begin{equation*}
q_0(t) \geq 0.
\end{equation*}
for any $t \in [0, T]$.
\end{document}